\documentclass[aps,prd,nofootinbib,11pt]{revtex4-1}
\usepackage{latexsym,bm,amsmath,amssymb,amsfonts}

\usepackage{graphicx,shortvrb}
\usepackage{epsfig}
\usepackage{newlfont}
\usepackage{hyperref}
\usepackage{xcolor}
\usepackage{cool}
\usepackage{booktabs}
\hypersetup{
    colorlinks,
    linkcolor={red!50!black},
    citecolor={blue!50!black},
    urlcolor={blue!80!black}
}

\newcommand{\ra}[1]{\renewcommand{\arraystretch}{#1}}

\AtBeginDocument{
\heavyrulewidth=.08em
\lightrulewidth=.05em
\cmidrulewidth=.03em
\belowrulesep=.65ex
\belowbottomsep=0pt
\aboverulesep=.4ex
\abovetopsep=0pt
\cmidrulesep=\doublerulesep
\cmidrulekern=.5em
\defaultaddspace=.5em
}
\begin{document}

\title{Circularly symmetric solutions of Minimal Massive Gravity at its merger point}

\author{{\"O}zg{\"u}r Sar{\i}o\u{g}lu}
\email{sarioglu@metu.edu.tr}
\affiliation{Department of Physics, Faculty of Arts and  Sciences,\\
              Middle East Technical University, 06800, Ankara, Turkey}

\date{\today}

\begin{abstract}
I find all the static circularly symmetric solutions of Minimal Massive 3D Gravity at 
its merger point, construct stationary versions of these and discuss some of their 
geometric and physical properties. It turns out that apart from a static hairy black hole,
there is also a static gravitational soliton, that has been overlooked in the literature.
\end{abstract}

\maketitle

\section{Introduction}\label{intro}
Since its inception, the AdS/CFT correspondence has been the game in town; not the 
only one for sure, but certainly the most prominent one. On the conformal field theory 
(CFT) side, a lot is known and well-understood about the two-dimensional CFTs, but it 
is well-known that on the gravity side the first go to guy, i.e. the three-dimensional 
cosmological Einstein's theory, has no propagating degrees of freedom and one is 
forced to look into the massive gravity theories. 

The cosmological Topologically Massive Gravity (TMG) \cite{Deser:1982vy}, where 
the ``wrong-sign"  Einstein-Hilbert action with a cosmological constant is paired up with 
the parity-odd gravitational Chern-Simons term, is the most seasoned one around, and 
indeed has a a single massive graviton mode around its convenient anti de-Sitter (AdS) 
vacuum. Unfortunately though, the central charge of the dual CFT is positive only if the 
bulk graviton has negative energy, which, of course, ruins unitarity. In a hope to fix this
``bulk vs boundary clash'' issue, the parity-preserving New Massive Gravity (NMG) 
model \cite{Bergshoeff:2009hq} was introduced. This theory is characterized by an action 
that consists of the addition of a particular combination of quadratic curvature terms to the 
usual (cosmological) Einstein-Hilbert term, but also suffers from the pathologies of TMG. 
A slightly spiced-up version of NMG, which is obtained by concatenating its action with 
the gravitational Chern-Simons term, yields the so-called Generalized Massive Gravity 
(GMG) in the menu. Finally, the delicacy is given by an extension of TMG, dubbed Minimal 
Massive Gravity (MMG) \cite{Bergshoeff:2014pca}, which remarkably overcomes the 
``bulk vs boundary clash" problem for a certain range of its parameters 
\cite{Arvanitakis:2014xna}. Even though MMG, just like its older sister TMG, has only 
one massive spin-2 mode in the bulk, its action is best described in a ``Chern-Simons-like 
formulation" using auxiliary fields and cannot be reduced to an action for the metric 
alone (see \cite{Bergshoeff:2014pca} for details). This makes it a nontrivial problem to 
extend MMG with matter couplings, which has been solved by employing a particular 
source tensor that is quadratic in the stress tensor \cite{Arvanitakis:2014yja}.

The possibility that ``unitary" MMG is a sensible candidate for a holographic theory of,
albeit massive, gravity makes it worth the effort to determine and study its exact solutions.
The celebrated BTZ black hole \cite{Banados:1992wn}, being locally AdS, ubiquitously
solves all the models described in the previous paragraph. A static ``hairy" black hole, 
as well as an (A)dS$_2\times$S$^1$ vacuum, a warped (A)dS black hole
\cite{Arvanitakis:2014yja}, a wave solution \cite{Alishahiha:2014dma}, a two-parameter 
deformation of the BTZ black hole \cite{Giribet:2014wla}, a non-BTZ AdS black 
hole \cite{Arvanitakis:2015yya}, Kundt solutions \cite{Deger:2015wpa} and (simply
transitive) homogeneous solutions \cite{Charyyev:2017uuu} are exact solutions that have 
been found thus far.

Among the works briefly alluded to in the previous paragraph, \cite{Arvanitakis:2015yya} 
deserves a special mention since it was shown in \cite{Arvanitakis:2015yya} that 
\emph{away from} its so-called \emph{merger point}, all conformally flat solutions 
of MMG must necessarily be (locally) maximally symmetric. Briefly stated, the merger 
point of MMG is the special point in the parameter space where the two possible 
values of the cosmological constant of a maximally symmetric solution coalesce 
\cite{Arvanitakis:2014yja}. Thus, in this work I will be working only at the merger point 
of MMG and, to narrow down the solution set, will only consider solutions that are 
\emph{circularly symmetric}. It will be shown that apart from the static ``hairy" black hole 
\cite{Arvanitakis:2014yja}, there is yet another static circularly symmetric solution of MMG 
at its merger point. Although this solution was already known in the context of NMG and 
interpreted as describing a gravitational soliton, no one, to my knowledge, has mentioned 
that it also solves MMG. These static solutions can literally be boosted to make them 
stationary, and again these rotating versions naturally solve MMG at its merger point.
The stationary black hole thus obtained can be identified with an already familiar solution 
of NMG, but the other one seems to be new. 

The organization of the paper is as follows: In section \ref{model}, I describe the massive 
gravity models briefly mentioned above, and then find the respective merger points for the
relevant ones. Section \ref{static} is devoted to a careful derivation of the static circularly 
symmetric solutions of MMG at its merger point, and a brief discussion of their elementary 
properties. Section \ref{stationary} presents the stationary versions of the static spacetimes 
found in section \ref{static}. I study the conserved gravitational charges of these solutions
in section \ref{charges}, and work out the entropy and thus the first law of thermodynamics 
for the black holes in section \ref{thermo}. I then conclude with a discussion of possible
future work. The outline of the calculation that takes one from the static black hole to its
stationary version is given in appendix \ref{appa}.

\section{The massive gravity models}\label{model}
I start with a review of the main pillars of three-dimensional massive gravity theories, which were
already alluded to in section \ref{intro}. Along the way, I clearly set my conventions and 
show the relations between these models. Then I explicitly present the merger points of the 
relevant ones.

\subsection{The TMG, MMG, NMG and GMG models}
The source-free field equation of TMG \cite{Deser:1982vy} is\footnote{With 
the conventions I use in this paper, one must in fact put a minus sign in front of the Einstein
tensor $G_{\mu\nu}$ in (\ref{TMG}) and accordingly the Ricci scalar $R$ in (\ref{tmg}) to 
take care of the ``wrong sign" issue of TMG.}
\begin{equation}
G_{\mu\nu} + \Lambda_{0} \, g_{\mu\nu} + \frac{1}{\mu} C_{\mu\nu} = 0 \,, \qquad 
G_{\mu\nu} \equiv R_{\mu\nu} - \frac{1}{2} R \, g_{\mu\nu} \,,
\label{TMG}
\end{equation}
which follows from the variation of the action
\begin{eqnarray}
I_\mathrm{TMG} & = & \int d^3 x \, \sqrt{-g} \, \big( R - 2 \Lambda_{0} \big) 
+ I_\mathrm{GCS} \,, \label{tmg} \\
I_\mathrm{GCS} & \equiv & \frac{1}{2\mu} \int d^3 x \, \sqrt{-g} \, \epsilon^{\sigma\mu\nu} \, 
\Gamma^{\rho}\,_{\sigma\tau} \, \Big( \partial_{\mu} \, \Gamma^{\tau}\,_{\rho\nu} 
+ \frac{2}{3} \, \Gamma^{\tau}\,_{\mu\lambda} \, \Gamma^{\lambda}\,_{\nu\rho} \Big) \,,
\label{gcs}
\end{eqnarray}
obtained by adding the gravitational Chern-Simons term $I_\mathrm{GCS}$ to the usual cosmological 
Einstein-Hilbert piece. Here the symmetric, traceless, parity-odd and covariantly conserved 
Cotton tensor $C_{\mu\nu}$ is defined in terms of the Schouten tensor $S_{\sigma\nu}$ as
\begin{equation}
S_{\sigma\nu} \equiv R_{\sigma\nu} - \frac{1}{4} R \, g_{\sigma\nu}  \,, \qquad 
C^{\mu}\,_{\nu} \equiv \epsilon^{\mu\rho\sigma} \, \nabla_{\rho} \, S_{\sigma\nu} \,, \qquad 
S \equiv g^{\mu\nu} S_{\mu\nu} = \frac{R}{4} \,,
\label{cotten}
\end{equation}
and the Levi-Civita pseudo tensor is defined as
\( \epsilon_{\mu\rho\sigma} = \sqrt{-g} \, \varepsilon_{\mu\rho\sigma} \)
in terms of the weight +1 tensor density \( \varepsilon_{\mu\rho\sigma} \), where I use 
the convention \( \varepsilon_{012} = +1 \). Here $\Lambda_{0}$ is the cosmological constant
(with dimensions 1/Length$^2$), and $\mu$ is a mass parameter (with dimensions 1/Length). 

The field equation of source-free MMG theory reads
\begin{equation}
E_{\mu\nu} \equiv \bar{\sigma} G_{\mu\nu} + \bar{\Lambda}_{0} \, g_{\mu\nu} 
+ \frac{1}{\mu} C_{\mu\nu} + \frac{\gamma}{\mu^2} J_{\mu\nu} = 0 \,,
\label{MMG}
\end{equation}
where the symmetric, curvature-squared tensor $J_{\mu\nu}$ is
\begin{eqnarray}
J^{\mu\nu} & \equiv & \frac{1}{2} \, \epsilon^{\mu\rho\sigma} \, \epsilon^{\nu\tau\eta} \, 
S_{\rho\tau} \, S_{\sigma\eta} = S S^{\mu\nu} - S^{\mu\rho} S^{\nu}\,_{\rho}
+ \frac{1}{2} g^{\mu\nu} \Big( S^{\rho\sigma} S_{\rho\sigma} - S^2 \Big)
\,, \label{jten} \\
J & \equiv & g_{\mu\nu} J^{\mu\nu} = \frac{1}{2} \Big( S^{\rho\sigma} S_{\rho\sigma} - S^2 \Big) 
\,. \nonumber 
\end{eqnarray}
Even though the tensor $J^{\mu\nu}$ is not covariantly conserved on its own, i.e. the 
MMG field equation (\ref{MMG}) can not be derived from an action that only involves the 
metric and its curvature tensors, it is nevertheless conserved 
\emph{on-shell} as a consequence of (\ref{MMG}) itself (see 
\cite{Bergshoeff:2014pca, Arvanitakis:2014yja} for details). In (\ref{MMG}), the parameters 
$\bar{\sigma}$ and $\gamma$ are non-zero dimensionless constants, whereas 
$\bar{\Lambda}_{0}$ is the bare cosmological constant. 

As a sister to the $J$-tensor, one can also consider the following symmetric, traceless,
curvature-squared tensor \cite{Afshar:2014ffa}
\[ H^{\mu\nu} \equiv \frac{1}{2} \, \epsilon^{\mu\rho\sigma} \nabla_{\rho} \, C^{\nu}\,_{\sigma}
+ \frac{1}{2} \, \epsilon^{\nu\rho\sigma} \nabla_{\rho} \, C^{\mu}\,_{\sigma} =
\square S^{\mu\nu} - \nabla^{\mu} \nabla^{\nu} S - 3 S^{\mu\rho} S^{\nu}\,_{\rho} 
+ g^{\mu\nu} S^{\rho\sigma} S_{\rho\sigma} \,, \]
which has the covariant derivative
\( \nabla_{\mu} H^{\mu\nu} = - \nabla_{\mu} J^{\mu\nu} = 
\epsilon^{\nu\tau\mu} \, C^{\sigma}\,_{\mu} \, S_{\sigma\tau} \). 
Denoting the sum of these tensors by \( K^{\mu\nu} \equiv J^{\mu\nu} + H^{\mu\nu} \) 
\cite{Afshar:2014ffa}, one has a Bianchi identity \( \nabla_{\mu} K^{\mu\nu} = 0 \), which
implies that $K^{\mu\nu}$ follows from the variation of an action. This is indeed the case
\cite{Afshar:2014ffa}, and one arrives at the source-free NMG action 
\cite{Bergshoeff:2009hq}
\begin{equation}
I_\mathrm{NMG} = \int d^3 x \, \sqrt{-g} \, \Big( \sigma R - 2 \Lambda_{0} + 
\frac{1}{m^2} \big( S^{\mu\nu} S_{\mu\nu} - S^2 \big) \Big) \,, \label{nmg}
\end{equation}
with the field equation
\begin{equation}
\sigma G_{\mu\nu} + \Lambda_{0} \, g_{\mu\nu} + \frac{1}{m^2} K_{\mu\nu} = 0 \,, \label{NMG}
\end{equation} 
where \( \sigma = \pm 1 \), $m$ is another mass parameter (with dimensions 1/Length) and
\[ K_{\mu\nu} = \square S_{\mu\nu} - \nabla_{\mu} \nabla_{\nu} S 
+ S S_{\mu\nu} - 4 S_{\mu\rho} S_{\nu}\,^{\rho} + \frac{1}{2} g_{\mu\nu} \Big( 3 S^{\rho\sigma} S_{\rho\sigma} - S^2 \Big) \,, \quad
K \equiv g^{\mu\nu} K_{\mu\nu} = J \,. \]

Finally, the sum of the $I_\mathrm{GCS}$ (\ref{gcs}) and the $I_\mathrm{NMG}$ 
(\ref{nmg}) terms, \( I_\mathrm{GMG} \equiv I_\mathrm{NMG} + I_\mathrm{GCS} \),
gives the so-called ``General Massive Gravity" (GMG) model 
\cite{Bergshoeff:2009hq} with the field equation
\begin{equation}
\sigma G_{\mu\nu} + \Lambda_{0} \, g_{\mu\nu} + \frac{1}{m^2} K_{\mu\nu} 
+ \frac{1}{\mu} C_{\mu\nu} = 0 \,. \label{GMG}
\end{equation} 

\subsection{The merger points}
For an Einstein space, \( R_{\mu\nu} = 2 \Lambda \, g_{\mu\nu} \)
(or \( S_{\mu\nu} = \tfrac{\Lambda}{2} g_{\mu\nu} \)), TMG (\ref{TMG}) has a unique vacuum given 
by\footnote{\( \Lambda = - \Lambda_{0} \), if one is to insert a minus sign in front of $G_{\mu\nu}$ 
in (\ref{TMG}).} \( \Lambda = \Lambda_{0} \), whereas MMG field equations  (\ref{MMG}) lead to 
the quadratic equation \cite{Bergshoeff:2014pca}
\begin{equation}
\Lambda^2 + \frac{4 \mu^2 \bar{\sigma}}{\gamma} \Lambda 
- \frac{4 \mu^2}{\gamma} \bar{\Lambda}_{0} = 0 \,, \label{lamMMG}
\end{equation}
which determines the two possible values of $\Lambda$ as
\begin{equation}
\Lambda_{\pm} = - \frac{2 \mu^2 \bar{\sigma}}{\gamma} \left( 1 \pm 
\sqrt{1+ \frac{\gamma \bar{\Lambda}_{0}}{\mu^2 \bar{\sigma}^{2}}} \, \right) \,, \label{lamval}
\end{equation}
provided the inequality \( \mu^2 \bar{\sigma}^{2} + \gamma \bar{\Lambda}_{0} \geq 0 \) holds. As for
NMG (\ref{NMG}) and GMG (\ref{GMG}), one arrives analogously at
\begin{equation}
\Lambda^2 + 4 m^2 \sigma \Lambda - 4 m^2 \Lambda_{0} = 0 \,, \qquad \mbox{and} \qquad
\Lambda_{\pm} = -2 m^2 \left( \sigma \pm \sqrt{1+ \frac{\Lambda_{0}}{m^2}} \right) \,, \label{lamNMG}
\end{equation}
provided \( m^2 + \Lambda_{0} \geq 0 \). As advertised earlier, in this work I will be working 
only at the so-called merger point, the special point in the parameter space of a given 
quadratic-curvature theory where the two possible values of $\Lambda$ coalesce. Thus, 
the merger point for MMG is
\begin{equation}
\bar{\Lambda}_{0} = - \frac{\mu^2 \bar{\sigma}^{2}}{\gamma} \,, \;\;
\Lambda = - \frac{2 \mu^2 \bar{\sigma}}{\gamma} \quad \mbox{and} \quad 
\bar{\sigma} \Lambda = 2 \bar{\Lambda}_{0} \,, \label{merMMG}
\end{equation}
whereas the merger point for NMG and GMG is
\begin{equation}
\Lambda_{0} = -m^2 \,, \;\; \Lambda = - 2 m^2 \sigma \quad \mbox{and} \quad
\sigma \Lambda = 2 \Lambda_{0} \,. \label{merNMG}
\end{equation}
Note that if $\sigma=1$, NMG and GMG has an AdS vacuum, and if $\sigma=-1$, then a dS vacuum.

\section{Static circularly symmetric solutions and their properties}\label{static}
Here I first carefully derive all the static circularly symmetric solutions of MMG at its merger
point, and then briefly state the most elementary properties of these.

\subsection{Static circularly symmetric solutions}\label{scss}
Let me consider the most general static, circularly symmetric metric
\begin{equation} 
ds^2 = - u(r) \, dt^2 + \frac{dr^2}{v(r)} + r^2 \, d\theta^2 \,, \label{ans}
\end{equation}
and find out all metrics of the form (\ref{ans}) that satisfy \( E_{\mu\nu} = 0 \) (\ref{MMG}) at the 
merger point (\ref{merMMG}). One quick observation to note is that all components of the 
Cotton tensor, apart from \( C_{t\theta} \), vanish for this metric, whereas only the $tt$-, $rr$- and 
$\theta\theta$-components of the remaining terms in $E_{\mu\nu}$ are non-vanishing. 
Moreover, it turns out that both the \( E_{rr} = 0 \) and \( E_{\theta\theta} = 0 \)
equations can separately be written as the product of two nontrivial factors, where one of 
each pair is common. Sparing the reader, this is best displayed by considering the linear
combination \( E^{r}\,_{r} + E^{\theta}\,_{\theta} = 0 \), which implies that
\begin{equation}
\big( 2 \Lambda r + v^{\prime} \big) \Big( 2 u^2 \big( 2 \Lambda r - v^{\prime} \big) 
+ r u^{\prime} \big( u v^{\prime} - v u^{\prime} \big) + 2 u v \big(r u^{\prime} \big)^{\prime} \Big) = 0 \,, 
\label{cases}
\end{equation}
where a prime indicates differentiation with respect to the coordinate $r$. Let me solve (\ref{cases})
by considering the vanishing of each factor on the left hand side separately.

\textbullet \(\; 2 \Lambda r + v^{\prime} = 0 :\) 
In this case one can quickly determine $v$ as \( v(r) = v_0 - \Lambda r^2 \) for an integration 
constant $v_0$. Using this, one finds from \( E_{tt} = 0 \) that the metric function $u$ must satisfy 
\[ r (v_0 - \Lambda r^2) \big( (u^{\prime})^2 - 2 u u^{\prime\prime} \big) + 2 v_0 u u^{\prime} = 0 
\quad \implies \quad u(r) = u_2 \big( \sqrt{v_0 - \Lambda r^2} - u_1 \big)^2 \,, \]
for integration constants $u_1$ and $u_2$. After a rescaling of the $t$-coordinate and a renaming 
of the integration constants, one arrives at the following metric:
\begin{equation}
ds_1^2 = \Lambda \left( - r_1 + \sqrt{r^2 - r_0} \, \right)^2 dt^2 - \frac{dr^2}{\Lambda (r^2 - r_0)} 
+ r^2 d\theta^2 \,. \label{met1}
\end{equation}

\textbullet \( \; 2 u^2 \big( 2 \Lambda r - v^{\prime} \big) 
+ r u^{\prime} \big( u v^{\prime} - v u^{\prime} \big) + 2 u v \big(r u^{\prime} \big)^{\prime}  = 0 :\)
Solving this equation algebraically for $u^{\prime\prime}$ and substituting it back into 
\( E_{tt} = 0 \) gives 
\( \big( u/v \big)^{\prime} \big( 2 \Lambda r u + v u^{\prime} \big) = 0, \)
with two new cases to study:

{\rm i)} \( \; (u/v)^{\prime} = 0 :\)
One can, without loss of generality, set $u=v$ in this case. Then the algebraic equation for 
$u^{\prime\prime}$ simplifies to \( u^{\prime\prime} = -2 \Lambda \), which is readily solved
as \( u(r) = v(r) = - \Lambda r^2 + u_1 r + u_2 \) for integration constants $u_1$ and $u_2$.
After a renaming of these, one arrives at 
\begin{equation}
ds_2^2 = \Lambda (r-r_{-}) (r-r_{+}) dt^2 - \frac{dr^2}{\Lambda (r-r_{-}) (r-r_{+})} + r^2 d\theta^2 \,.
\label{met2} 
\end{equation}

{\rm ii)}  \( \; 2 \Lambda r u + v u^{\prime} = 0 :\)
Solving this constraint for $u^{\prime}$ and demanding that it be compatible with the algebraic 
equation for $u^{\prime\prime}$, one arrives at 
\( (v - \Lambda r^2) \big( 2 \Lambda r + v^{\prime} \big) = 0 \). 
Considering the analysis of the \( 2 \Lambda r + v^{\prime} = 0 \) case above, one finds that 
demanding the second factor to vanish leads one to the metric (\ref{met1}) with the integration 
constant $r_1=0$, which can also be identified with the \( r_{-} + r_{+} = 0 \) form of (\ref{met2}). 
Thus, there is no new solution per se. Finally, setting \( v(r) = \Lambda r^2 \), one finds 
\( u(r) = u_0/r^2 \), for an integration constant $u_0$. Taking $u_0=1$, one arrives at the metric
\begin{equation}
ds_\mathrm{L}^2 = - \frac{dt^2}{r^2} + \frac{dr^2}{\Lambda r^2} + r^2 d\theta^2 \,,
\label{lif}
\end{equation}
which is easily identified as the static Lifshitz spacetime \cite{Kachru:2008yh} with the 
dynamical exponent $z=-1$ \cite{Charyyev:2017uuu} when $\Lambda > 0$. However, a close 
scrutiny immediately reveals that  \( C_{t\theta} = 2 \Lambda^{3/2} \neq 0 \), so that 
the field equation for MMG (\ref{MMG}) is \emph{not} satisfied by (\ref{lif}).

This exhausts all the possibilities to consider. I have shown that (\ref{met1}) and (\ref{met2}) 
are the only static circularly symmetric solutions of MMG (\ref{MMG}) at the merger point
(\ref{merMMG}). Note, however, that for both solutions the $C_{\mu\nu}$ and, 
hence, $H_{\mu\nu}$ vanish identically: Given that one is a solution to either NMG, 
MMG or GMG, it is immediate to see that it is also a solution to the others. 

The solution (\ref{met2}) is long known and well understood 
\cite{Bergshoeff:2009hq,Arvanitakis:2014yja,Oliva:2009ip}, whereas the version of (\ref{met1})
for $r_0<0$, even though known in the context of NMG \cite{Oliva:2009ip,Perez:2011qp}, has 
been comparatively less studied. The spacetime (\ref{met2}) was shown to be a solution of 
MMG at the merger point in \cite{Arvanitakis:2014yja}, but no one, to my knowledge, mentioned 
that (\ref{met1}) is also such a solution. 

\subsection{The basic properties of the static solutions}
For convenience, I briefly list some of the properties of (\ref{met2}) here and 
refer the reader to \cite{Bergshoeff:2009hq,Arvanitakis:2014yja,Oliva:2009ip} for details.
The curvature invariants read
\begin{equation}
R = 6 \Lambda -\frac{2 \Lambda (r_{-} + r_{+})}{r} \,, \quad
R^{\mu\nu} R_{\mu\nu} = 12 \Lambda^2 
+ \frac{\Lambda^2 (r_{-} + r_{+}) \big(3(r_{-} + r_{+})-16 r \big)}{2 r^2} \,,
\label{rick2}
\end{equation}
and clearly (\ref{met2}) is not a space of maximum symmetry even though it asymptotically 
becomes one as $r \to \infty$. With the assumption that \( r_{+} \geq r_{-} \), one has an 
asymptotically AdS black hole with an event horizon at \( r = r_{+} \) for $\Lambda < 0$. 
The near horizon geometry of the zero-temperature extremal black hole (given by 
\( r_{+} = r_{-} \)) in this case is AdS$_2 \times$S$^1$. When the ``hair" 
\( r_{-} + r_{+} = 0 \), the metric (\ref{met2}) can be shown to be identical to the metric 
of the static BTZ black hole \cite{Banados:1992wn}. As for $\Lambda > 0$, \( r = r_{+} \) 
is the cosmological horizon and if $r_{-} > 0$, then there is a black hole with horizon 
at \( r = r_{-} \) in a static dS vacuum for \( r_{+} > r > r_{-} \). In the limit \( r_{-} \to r_{+} \), 
the geometry becomes dS$_2 \times$S$^1$. 

On the other hand, it is clear that the solution (\ref{met1}) is meaningful only 
if $r^2 > r_0$. The curvature invariants read
\begin{equation}
R = 6 \Lambda + \frac{4 \Lambda r_1}{\big(- r_1 + \sqrt{r^2 - r_0} \, \big)} \quad \mbox{and} \quad
R^{\mu\nu} R_{\mu\nu} = 
4 \Lambda^2 + \frac{2 \Lambda^2 (4 r^2 - 4 r_0 - r_1^2)}{\big(- r_1 + \sqrt{r^2 - r_0} \, \big)^2} \,. 
\label{rick1}
\end{equation}
If the ``hair" $r_1 = 0$, then (\ref{met1}) describes a space of maximal symmetry and, as is the 
case for (\ref{met2}) when \( r_{-} + r_{+} = 0 \), it can be identified with the metric of the ``static" 
BTZ black hole when $\Lambda < 0$. When $r_1 < 0$, there is no singularity and the metric is
regular everywhere. However, when $r_1 > 0$ and $r_0 \geq - r_1^2$, there is a naked 
singularity at a finite $r$ given by the root of the $tt$-component of the metric, located at 
\( r = r_* = \sqrt{r_0 + r_1^2} \geq 0 \). As $r \to \infty$, (\ref{met1}) asymptotically approaches 
a space of maximal symmetry. If one naively takes the ``hairless" limit \( r_{-} + r_{+} \to 0 \) 
in (\ref{met2}), one obtains (\ref{met1}) with no ``hair"  $r_1 = 0$, which is locally (A)dS. 

The coordinate transformation \( r = \sqrt{r_0} \, \cosh{x} \) (assuming $r_0 > 0$) and a 
suitable scaling of the $t$ and $\theta$ coordinates take (\ref{met1}) to the form 
\begin{equation}
ds_1^2 = \frac{1}{\Lambda} \left( \big( x_0 + \sinh{x} \big)^2 \, dt^2 - dx^2 
+ \Lambda \cosh^2{x} \, d\theta^2 \right) \,, \label{stmet1a}
\end{equation}
with a single parameter \( x_0 \equiv -r_1/\sqrt{r_0} \). The curvature scalar of 
(\ref{stmet1a}) is
\[ R = 2 \Lambda + \frac{4 \Lambda \sinh{x}}{x_0 + \sinh{x}} \,, \]
and this spacetime is regular everywhere provided \( x_0 + \sinh{x} > 0 \). For the case $r_0 < 0$,
one can similarly employ the transformation \( r = \sqrt{-r_0} \, \sinh{y} \) to arrive at
\begin{equation}
ds_1^2 = \frac{1}{\Lambda} \left( \big( y_0 + \cosh{y} \big)^2 \, dt^2 - dy^2 
+ \Lambda \sinh^2{y} \, d\theta^2 \right) \,, \label{stmet1b}
\end{equation}
with a parameter \( y_0 \equiv -r_1/\sqrt{-r_0} \). Analogously, one finds
\[ R = 2 \Lambda + \frac{4 \Lambda \cosh{y}}{y_0 + \cosh{y}} \,, \]
and this is also regular everywhere provided \( y_0 + \cosh{y} > 0 \). According to
\cite{Oliva:2009ip}, the latter (\ref{stmet1b}) can be interpreted as a gravitational soliton (with 
a negative fixed mass) and the constant $y_0$ can be thought of as a modulus parameter 
(see also \cite{Perez:2011qp} for details).

\section{Stationary solutions}\label{stationary}
Since the definition of conserved gravitational charges using Killing isometries is problematic 
by the very existence of a cosmological horizon for a dS background, I will mostly be concerned 
with the versions of NMG, MMG and GMG theories that admit an AdS vacuum from now on. 
Thus, unless kept explicitly, I set $\Lambda = -1/\ell^2 < 0$ for the rest of this work.

\subsection{Stationary black hole}\label{sbh}
The stationary spacetime 
\begin{eqnarray}
d\tilde{s}_2^2 & = & \big( r^2 - (1+\omega^2) f(r) \big) \frac{dt^2}{\omega^2 \ell^2} 
+ \frac{2 dt d\theta}{\omega \ell} \big( f(r) - r^2 \big) + r^2 d\theta^2 
+ \frac{\omega^2 \, \ell^2 \, r^2 \, dr^2}{(r^2+\alpha^2-\beta) \big( f(r) - (1-\omega^2) r^2 \big)} 
\,, \nonumber \\
f(r) & = & \big( \alpha \pm \sqrt{r^2+ \alpha^2-\beta} \, \big)^2 \,, \label{rmet2} 
\end{eqnarray}
where $\omega$ is a new dimensionless parameter with \( 0 < | \omega | < 1 \), is a solution 
to the MMG, NMG and GMG theories at their respective merger points. This solution has been
obtained from (\ref{met2}) by ``boosting" it\footnote{See appendix \ref{appa} for an outline of 
this calculation.}, an old trick which has been employed in the three-dimensional gravity 
context for a long time (see e.g. \cite{Martinez:1999qi} for its relevance to the celebrated 
BTZ metric). As it stands, the parameters $\alpha$ and $\beta$ in (\ref{rmet2}) are related to 
the original $r_{-}$ and $r_{+}$ in (\ref{met2}) as
\begin{equation} 
\alpha \equiv \frac{\omega^2 \, (r_{-} + r_{+})}{2 (1-\omega^2)} 
\quad \mbox{and} \quad \beta \equiv - \frac{\omega^2 \, r_{-}  r_{+}}{(1-\omega^2)} \,, \label{albe}
\end{equation}
but one can completely forget about this and view these as completely independent of their parents,
as part of an autonomous solution. Note that (\ref{rmet2}) can also be written as
\begin{equation}
d\tilde{s}_2^2 = - \frac{f(r)\big( f(r) - (1-\omega^2) r^2 \big)}{\omega^2 \ell^2 r^2} dt^2 
+ \frac{\omega^2 \ell^2 r^2 \, dr^2}{(r^2+\alpha^2-\beta) \big( f(r) - (1-\omega^2) r^2 \big)} 
+ r^2 \Big( d\theta + \frac{f(r) - r^2}{\omega \ell r^2} dt \Big)^2 \,, \label{altrmet2}
\end{equation}
which is also convenient. 

As stated implicitly, the Cotton tensor \(C_{\mu\nu}=0 \) identically for this metric just like its seed (\ref{met2}). The curvature invariants of (\ref{rmet2}) are
\begin{equation}
R = - \frac{6}{\ell^2} - \frac{4 \alpha (1 - \omega^2)}{\ell^2 \omega^2 \sqrt{f(r)}} \,, \quad
R^{\mu\nu} R_{\mu\nu} = \frac{12}{\ell^4} + \frac{\alpha (1 - \omega^2)}{\ell^4 \omega^2 f(r)}
\Big( \frac{6 \alpha (1 - \omega^2)}{\omega^2} +16 \sqrt{f(r)} \, \Big) \,.
\label{rotrick2}
\end{equation}
Since (\ref{rmet2}) is left invariant under \( r \mapsto -r \), it is enough to consider \( r \geq 0 \) only.
It quickly follows that (\ref{rmet2}) describes a black hole with a singularity at \(f(r) = 0 \) or 
\( r^2 = \beta \geq 0 \) regardless of $\alpha$, provided one chooses the sign in the expression of 
$f(r)$ properly: If \( \alpha \geq 0 \), then one should choose the negative sign, and accordingly
the positive sign if \( \alpha < 0 \). To simplify the discussion, I will assume that 
\begin{equation} 
\beta \geq 0 \,, \qquad \alpha \geq 0  \qquad \mbox{and} \qquad
f(r) = \big( \alpha - \sqrt{r^2+ \alpha^2-\beta} \, \big)^2 
\label{choices}
\end{equation} 
for the remainder of this paper. Finally, the event horizon is located at the larger root 
of the equation \( f(r) = (1 - \omega^2) r^2 \), which is given explicitly as
\begin{equation}
r_{h} = \frac{\alpha \sqrt{1-\omega^2}}{\omega^2} 
\left( 1 + \sqrt{1 + \frac{\beta \omega^2}{\alpha^2 (1-\omega^2)}} \, \right) \,, \label{rhdef}
\end{equation}
Note that in the ``hairless" limit $\alpha \to 0$, one has \( r_{h} \to \sqrt{\beta} / \omega > \sqrt{\beta} \)
since by assumption \( 0 < | \omega | < 1 \). Moreover, (\ref{rmet2}) can be called as the 
\emph{``hairy" BTZ metric} since, when $\alpha \to 0$ (\ref{rmet2}) reduces to the canonical 
form of the BTZ metric (\ref{BTZ}) with the identifications
\[ M = \frac{\beta \, (1+\omega^2)}{\ell^2 \omega^2} \,, \qquad J = \frac{2 \beta}{\ell \omega} \,. \]

When one makes the following choice of the parameters  
\begin{equation}
\alpha = - \frac{a^2 b}{4 \Xi^{1/2} (1+\Xi^{1/2})} \,, \; 
\beta = \frac{a^4 b^2}{8 \Xi (1+\Xi^{1/2})^2} + \frac{a^2 M}{2 (1+\Xi^{1/2})} \,,\; 
\omega = \frac{a}{\ell (1+\Xi^{1/2})} \,, \; \Xi \equiv 1- \frac{a^2}{\ell^2} \label{ott}
\end{equation}
in (\ref{rmet2}), this solution becomes identical to the solution given in Section VI of \cite{Oliva:2009ip},
which I reproduce here (by setting their $G=1/4$) for later convenience
\begin{eqnarray}
ds_\mathrm{OTT}^2 & = & - N(r) \, F(r) \, dt^2 + r^2 \big( d\theta + K(r) \, dt \big)^2 + \frac{dr^2}{F(r)} \,,
\label{OTT} \\
H(r) & = & \Big( r^2 - \frac{M}{2} \ell^2 (1- \Xi^{1/2}) - \frac{b^2 \ell^4}{16} (1- \Xi^{-1/2})^2 \Big)^{1/2} \,,
\nonumber \\
F(r) & = & \frac{(H(r))^2}{r^2} \left( \frac{(H(r))^2}{\ell^2} + \frac{b}{2} (1+\Xi^{-1/2}) H(r) 
+ \frac{b^2 \ell^2}{16} (1- \Xi^{-1/2})^2 - M \Xi^{1/2} \right) \,, \nonumber \\
N(r) & = & \Big( 1 - \frac{b \ell^2}{4 H(r)} (1- \Xi^{-1/2}) \Big)^2 \,, \qquad 
K(r) = -\frac{a}{2 r^2} \big( M - b \Xi^{-1/2} H(r) \big) \,, \; \nonumber
\end{eqnarray}
with \( |a| < \ell \). So the three parameters $\alpha, \beta$ and $\omega$ of (\ref{rmet2}) are
related to the three parameters $a, b$ and $M$ of (\ref{OTT}) via (\ref{ott}). The
representation (\ref{rmet2}) certainly seems to be easier to handle compared to (\ref{OTT}).
Let me emphasize that no one, to my knowledge, has mentioned that (\ref{rmet2}) (or  (\ref{OTT})
for that matter) is a solution to MMG at its merger point.

\subsection{The other stationary solutions}
The ``boosting" trick can also be employed on the other solution (\ref{met1}) to 
find yet a second stationary metric that solves the MMG, NMG and GMG models 
at their merger points:
\begin{eqnarray}
d\tilde{s}_1^2 & = & \big( r^2 - (1+\omega^2) h(r) \big) \frac{dt^2}{\omega^2 \ell^2} 
+ \frac{2 dt d\theta}{\omega \ell} \big( h(r) - r^2 \big) + r^2 d\theta^2 
+ \frac{\ell^2 \, r^2 \, dr^2}{h(r)
\left( r^2 - \frac{R_1}{\omega^2} + \frac{(2\omega^2 -1)}{\omega^4} R_0^2 \right)} \,, \nonumber \\
h(r) & = & r^2 + 2 R_0^2 - R_1 \pm 2 R_0 \sqrt{r^2 - \frac{R_1}{\omega^2} 
+ \frac{(2\omega^2 -1)}{\omega^4} R_0^2 } \,. \label{rmet1}
\end{eqnarray}
Analogous to the relation mentioned between $(\alpha, \beta)$ and $(r_{-}, r_{+})$ pairs,
the new constants $R_0$ and $R_1$ are related to their parents $(r_0,r_1)$ as
\begin{equation} 
R_0 \equiv \frac{\omega^2}{(1-\omega^2)} r_1 \quad \mbox{and} \quad 
R_1 \equiv \frac{\omega^2}{(1-\omega^2)} \big( r_0 -r_1^2 \big) \,. \label{R0R1}
\end{equation}
Note that (\ref{rmet1}) can also be cast as
\begin{equation}
d\tilde{s}_1^2 = - \frac{h(r)\big( h(r) - (1-\omega^2) r^2 \big)}{\omega^2 \ell^2 r^2} dt^2 
+ \frac{\ell^2 \, r^2 \, dr^2}{h(r)
\left( r^2 - \frac{R_1}{\omega^2} + \frac{(2\omega^2 -1)}{\omega^4} R_0^2 \right)} 
+ r^2 \Big( d\theta + \frac{h(r) - r^2}{\omega \ell r^2} dt \Big)^2 \,. \label{altrmet1}
\end{equation}
When $R_0 \to 0$ (\ref{rmet1}) becomes identical to the BTZ black hole (\ref{BTZ}) with 
the identifications
\[ M = \frac{R_1 \, (1+\omega^2)}{\ell^2 \omega^2} \,, \qquad J = \frac{2 R_1}{\ell \omega} \,. \]

The Cotton tensor \( C_{\mu\nu} = 0 \) for this metric as well. For (\ref{rmet1}) to be
well-defined, it must be that
\[ r^2 > \tilde{r}^2 \equiv \frac{R_1}{\omega^2} + \frac{(1- 2 \omega^2) R_0^2}{\omega^4} \geq 0 \]
as a minimum requirement. On the other hand, the curvature scalar of (\ref{rmet1}) is
\[ R = -\frac{6}{\ell^2} - \frac{2 (1- \omega^2)}{\ell^2} 
\left( \frac{2 R_0^2 (1+ \omega^2) -\omega^2 \big( h(r)-r^2+R_1 \big)}{r^2 \omega^4 - 2 R_0^2 (1- \omega^2) -\omega^2 R_1} \right) \,, \]
which shows that $R$ diverges when 
\( r^2 \omega^4 - 2 R_0^2 (1- \omega^2) -\omega^2 R_1 = 0 \), i.e.
\[ r^2 = \tilde{r}^2 + \frac{R_0^2}{\omega^4} \,. \]
Thus, provided that \( \tilde{r}^2 \geq 0 \), the metric (\ref{rmet1}) is regular everywhere. As is shown
in section \ref{charges}, this solution has finite gravitational charges and can be thought of
as a rotating soliton.

\section{Gravitational charges}\label{charges}
Let me briefly summarize how gravitational charges are defined by making use of  
the background Killing isometries. (I refer the reader to \cite{Deser:2002jk} for details.) 
Let $h_{\mu\nu}$ denote the deviation of an asymptotically AdS metric from an AdS 
background, i.e. \( h_{\mu\nu} \equiv g_{\mu\nu} - \bar{g}_{\mu\nu} \) and $\bar{g}_{\mu\nu}$ 
satisfies \( \bar{R}_{\mu\nu} = 2 \Lambda \bar{g}_{\mu\nu} \) and \( \bar{R} = 6 \Lambda \),
with $\Lambda < 0$. Let me also assume that the deviation $ h_{\mu\nu} $ goes to zero sufficiently
fast as one approaches the background $\bar{g}_{\mu\nu}$, which is typically located at the
boundary at infinity. For a background Killing vector $\bar{\xi}^{\mu}$ that is smooth and well-defined 
in the geometry described by $\bar{g}_{\mu\nu}$, the following gives a conserved and background
gauge invariant gravitational charge\footnote{Here the constant $\kappa$ is proportional 
to the three-dimensional Newton's constant with dimensions of Length.}
\begin{equation}
Q^{\mu}(\bar{\xi}) = \frac{1}{\kappa} \oint_{\partial \Sigma} d\bar{S}_{\nu} \, {\cal F}^{\mu\nu} \,,
\label{charge}
\end{equation}
where ${\cal F}_{\mu\nu}$ is an antisymmetric tensor, $\partial \Sigma$ is a one-dimensional
boundary with the ``surface" element $d\bar{S}$. (In the ensuing discussion, this will be a line 
integral along a circle of radius $r \to \infty$.) For the conventions adopted in this work, one has
\begin{eqnarray}
{\cal F}^{\mu\nu}_\mathrm{Ein}(\bar{\xi}) & = & \bar{\xi}_{\sigma} \bar{\nabla}^{[\mu} h^{\nu]\sigma} 
+ \bar{\xi}^{[\nu} \bar{\nabla}_{\sigma} h^{\mu]\sigma} 
+ h^{\sigma[\nu} \bar{\nabla}_{\sigma} \bar{\xi}^{\mu]} + \bar{\xi}^{[\mu} \bar{\nabla}^{\nu]} h 
+ \frac{1}{2} h \bar{\nabla}^{[\mu} \bar{\xi}^{\nu]} \,, \label{einchar} \\
{\cal F}^{\mu\nu}_\mathrm{GCS} & = & \frac{1}{2\mu} \Big( {\cal F}^{\mu\nu}_\mathrm{Ein}(\bar{\Xi}) 
+ \bar{\epsilon}^{\sigma\nu\tau} \bar{\xi}_{\sigma} \big( S_{\tau}\,^{\mu} \big)_{L}
+ \bar{\epsilon}^{\mu\nu\tau} \bar{\xi}_{\sigma} \big( S_{\tau}\,^{\sigma} \big)_{L}
+ \bar{\epsilon}^{\mu\sigma\tau} \bar{\xi}_{\sigma} \big( S_{\tau}\,^{\nu} \big)_{L}
+ \bar{\epsilon}^{\mu\nu\sigma} \bar{\xi}_{\sigma}  S_{L} \Big) \,, \label{GCScharden} \\
{\cal F}^{\mu\nu}_\mathrm{NMG} & = & \Big( \sigma - \frac{\Lambda}{2m^2} \Big) 
{\cal F}^{\mu\nu}_\mathrm{Ein}(\bar{\xi}) + \frac{1}{m^2} \Big( 
2 \bar{\xi}^{\sigma} \bar{\nabla}^{[\nu} \big( S^{\mu]}\,_{\sigma}\big)_{L} 
+ 2 \big( S^{[\nu}\,_{\sigma}\big)_{L} \bar{\nabla}^{\mu]} \bar{\xi}^{\sigma} 
+ S_{L} \bar{\nabla}^{[\nu} \bar{\xi}^{\mu]} \Big) \,, \label{NMGcharden} \\
{\cal F}^{\mu\nu}_\mathrm{MMG} & = & \Big( \bar{\sigma} + \frac{\gamma \Lambda}{2 \mu^2} \Big) 
{\cal F}^{\mu\nu}_\mathrm{Ein}(\bar{\xi}) + {\cal F}^{\mu\nu}_\mathrm{GCS} \,, \label{MMGcharden} \\
{\cal F}^{\mu\nu}_\mathrm{GMG} & = & {\cal F}^{\mu\nu}_\mathrm{NMG} + 
{\cal F}^{\mu\nu}_\mathrm{GCS} \,, \label{GMGcharden}
\end{eqnarray}
where all contractions, raising and lowering of indices are performed with respect to
$\bar{g}_{\mu\nu}$, $\bar{\nabla}$ indicates the covariant derivative with respect to the 
background metric and the antisymmetrization of indices in the usual way is understood, e.g. 
\( A_{[\mu\nu]} \equiv \tfrac{1}{2} ( A_{\mu\nu} - A_{\nu\mu} ) \). Here 
\( \bar{\Xi}^{\mu} \equiv \bar{\epsilon}^{\mu\nu\sigma} \bar{\nabla}_{\nu} \bar{\xi}_{\sigma} \)
is another Killing vector constructed out of $\bar{\xi}$,
\begin{eqnarray}
\big( S^{\mu}\,_{\nu}\big)_{L} & = & \big( R^{\mu}\,_{\nu}\big)_{L} 
- \frac{1}{4} \delta^{\mu}\,_{\nu} \, R_{L} - 2 \Lambda h^{\mu}\,_{\nu} \,, \label{Schoutenlin} \\
R_{\mu\nu}^{L} & = & \frac{1}{2} \left( \bar{\nabla}^{\sigma} \bar{\nabla}_{\nu} h_{\mu\sigma} 
 + \bar{\nabla}^{\sigma} \bar{\nabla}_{\mu} h_{\nu\sigma} 
 - \bar{\square} h_{\mu\nu}
 - \bar{\nabla}_{\mu} \bar{\nabla}_{\nu} h \right) \,, \label{riclin} \\
R_{L} & = & 4 S_{L} = \bar{\nabla}^{\mu} \bar{\nabla}^{\nu} h_{\mu\nu} 
- \bar{\square} h - 2 \Lambda h  \,, \label{Schoutenscalin}
 \end{eqnarray}
\( h \equiv \bar{g}^{\mu\nu} h_{\mu\nu} \) and 
\( \bar{\square} \equiv \bar{\nabla}^{\mu} \bar{\nabla}_{\mu} \). Suffice it to say that 
all of these expressions, of course, reduce to their counterparts derived in \cite{Deser:2002jk}, 
but, wherever appropriate, are given in terms of the linearized Schouten tensor instead of the 
linearized cosmological-Einstein tensor \( {\cal G}_{\mu\nu}^{L} \).

Note that at the merger point (\ref{merNMG}) of NMG and GMG, the coefficient of 
${\cal F}^{\mu\nu}_\mathrm{Ein}(\bar{\xi})$ in (\ref{NMGcharden}) precisely equals 
$2 \sigma = 2$ for an AdS vacuum, whereas the analogous term in (\ref{MMGcharden}) 
vanishes identically at the merger point (\ref{merMMG}) of MMG. This implies that 
\( {\cal F}^{\mu\nu}_\mathrm{MMG} = {\cal F}^{\mu\nu}_\mathrm{GCS} \), and thus 
\( {\cal F}^{\mu\nu}_\mathrm{GMG} = {\cal F}^{\mu\nu}_\mathrm{NMG} 
+ {\cal F}^{\mu\nu}_\mathrm{MMG} \) from (\ref{GMGcharden}).
Hence, in what follows, I will focus on the calculation of the conserved charges in the NMG 
and MMG settings only. 

\subsection{The BTZ black hole} 
Let me determine the energy and angular momentum of the canonical BTZ metric 
\cite{Banados:1992wn}, given by 
\begin{equation} 
ds_\mathrm{BTZ}^2 = \Big( M - \frac{r^2}{\ell^2} \Big) \, dt^2 - J \, dt \, d\theta + r^2 \, d\theta^2
 + \frac{dr^2}{- M + \frac{r^2}{\ell^2} + \frac{J^2}{4 r^2}} \,, \label{BTZ}
\end{equation}
to demonstrate the gist of the calculation and to fix the calibration of the constant $\kappa$ 
in (\ref{charge}). Considering the fact that the BTZ metric solves the cosmological Einstein theory,
if one simply takes the background metric to be the AdS metric obtained by setting 
\( M=J=0 \) in (\ref{BTZ}), i.e. use (\ref{back}) below as background, 
chooses \( {\cal F}^{\mu\nu} = {\cal F}^{\mu\nu}_\mathrm{Ein}(\bar{\xi}) \) and employs first the 
timelike Killing vector \( \bar{\xi}^{\mu} = -\big( \partial/\partial t \big)^{\mu} \)
and then the spacelike Killing vector \( \bar{\xi}^{\mu} = \big( \partial/\partial \theta \big)^{\mu} \) 
in (\ref{charge}), one finds the energy and the angular momentum to be
\( E = \pi M/\kappa \) and \( L = \pi J/\kappa \), respectively. So for convenience, let me fix the 
constant $\kappa$ in (\ref{charge}) to be \( \kappa = \pi \) and do away with it completely to 
simplify the discussion. Thus for the BTZ metric, considered as a solution to the cosmological 
Einstein theory, one has $E=M$ and $L=J$ in terms of the parameters in the metric itself. 
An analogous calculation using \( {\cal F}^{\mu\nu} = {\cal F}^{\mu\nu}_\mathrm{NMG} \) 
(\ref{NMGcharden}) in (\ref{charge}) instead shows that 
\[ E_\mathrm{NMG} = 2 M \qquad \mbox{and} \qquad  L_\mathrm{NMG} = 2 J \] 
at the merger point (\ref{merNMG}) of NMG, whereas one has
\[ E_\mathrm{MMG} = \frac{J}{\mu \ell^2} \qquad \mbox{and} \qquad 
L_\mathrm{MMG} = \frac{M}{\mu} \,, \]
with an interchanging of the roles of the parameters $M$ and $J$ at the merger point 
(\ref{merMMG}) of MMG.

One can determine the energies of the static solutions (\ref{met1}) and (\ref{met2}), and similarly the
stationary solutions (\ref{rmet2}), (\ref{OTT}) and (\ref{rmet1}), at the respective merger 
points (\ref{merMMG}) and (\ref{merNMG}) of MMG and NMG in a similar fashion. 
Setting $\Lambda = -1/\ell^2$ as in the BTZ case, it is obvious that the background metric to 
work with (for both the static and stationary cases) is simply the AdS space in the Poincar\'{e} patch
\begin{equation} 
d\bar{s}^2 = - \frac{r^2}{\ell^2} \, dt^2 + \frac{\ell^2}{r^2} \, dr^2 + r^2 \, d\theta^2 \,, \label{back}
\end{equation}
obtained by setting \( r_{-} \to 0, r_{+} \to 0 \) in (\ref{met2}), \( r_{0} \to 0, r_{1} \to 0 \) in (\ref{met1}),
\(\alpha \to 0, \beta \to 0 \) in (\ref{rmet2}), \( M \to 0, b \to 0 \) in (\ref{OTT}) and 
\(R_0 \to 0, R_1 \to 0 \) in (\ref{rmet1}), respectively.

\subsection{The static solutions}
As in the BTZ case, the timelike Killing vector \( \bar{\xi}^{\mu} = -\big( \partial/\partial t \big)^{\mu} \)
can be employed to yield the energies of these solutions. Both (\ref{met1}) and (\ref{met2}) 
have $E_\mathrm{MMG} = 0$ in MMG, whereas (\ref{met2}) has energy 
\begin{equation}
E_\mathrm{NMG} = - \frac{2 r_{-} r_{+}}{\ell^2} \,, \label{Emet2NMG}
\end{equation}
(so that \(E_\mathrm{NMG} = 2 r^2_{+}/\ell^2 > 0\) in the ``hairless" limit \( r_{-} + r_{+} \to 0 \)) 
and (\ref{met1}) has energy 
\begin{equation}
E_\mathrm{NMG} = \frac{2 (r_{0} - r_{1}^2)}{\ell^2} \,, \label{Emet1NMG}
\end{equation}
(so that \( E_\mathrm{NMG} = 2 r_{0}/\ell^2 \) in the ``hairless" limit \( r_{1} \to 0 \)) in NMG. 

\subsection{The stationary solutions}
As for the stationary solutions, one analogously finds for the energies 
\begin{equation}
E_\mathrm{NMG} = \left\{
\begin{array}{ll}
-2 \frac{(2 \alpha^2 -\beta) (1+ \omega^2)}{\ell^2 \omega^2} & \mbox{for} \; (\ref{rmet2}), \\
-2 \frac{(2 R_0^2 - R_1) (1+ \omega^2)}{\ell^2 \omega^2} & \mbox{for} \; (\ref{rmet1}),
\end{array}\right. \label{ErmetNMG}
\end{equation}
in the NMG setting. Note that using (\ref{albe}), and respectively (\ref{R0R1}), in (\ref{ErmetNMG}) 
then taking the $\omega \to 0$ limit correctly yields (\ref{Emet2NMG}) and (\ref{Emet1NMG}), 
respectively, as it must be. On the other hand, in the MMG setting, one finds
\begin{equation}
E_\mathrm{MMG}  = \left\{
\begin{array}{ll}
-2 \frac{(2 \alpha^2 - \beta)}{\mu \omega \ell^3} & \mbox{for} \; (\ref{rmet2}),  \\
-2 \frac{(2 R_0^2 - R_1)}{\mu \omega \ell^3} &  \mbox{for} \; (\ref{rmet1}).
\end{array}\right. \label{ErmetMMG}
\end{equation}
Note again that using (\ref{albe}), and respectively (\ref{R0R1}), in (\ref{ErmetMMG}) 
and taking $\omega \to 0$ afterwards gives $E_\mathrm{MMG}=0$ for both cases, as it should. 
The angular momenta of the stationary solutions come by the spacelike Killing vector 
\( \bar{\xi}^{\mu} = \big( \partial/\partial \theta \big)^{\mu} \) as for the BTZ metric and read
\begin{equation}
L_\mathrm{NMG} = \left\{
\begin{array}{ll}
-4 \frac{(2 \alpha^2 -\beta)}{\omega \ell} & \mbox{for} \; (\ref{rmet2}), \\
-4 \frac{(2 R_0^2 - R_1)}{\omega \ell} & \mbox{for} \; (\ref{rmet1})
\end{array}\right. \label{LrmetNMG}
\end{equation}
in the NMG setting. Note that using (\ref{albe}), and respectively (\ref{R0R1}), in (\ref{LrmetNMG}) 
and taking $\omega \to 0$ later gives $L_\mathrm{NMG}=0$ for both cases, as expected. 
On the other hand, one finds
\begin{equation}
L_\mathrm{MMG} = \left\{
\begin{array}{ll}
- \frac{(2 \alpha^2 -\beta) (1+ \omega^2)}{\mu \omega^2 \ell^2} & \mbox{for} \; (\ref{rmet2}), \\
- \frac{(2 R_0^2 - R_1) (1+ \omega^2)}{\mu \omega^2 \ell^2} & \mbox{for} \; (\ref{rmet1})
\end{array}\right. \label{LrmetMMG}
\end{equation}
for the generic versions of both (\ref{rmet2}) and (\ref{rmet1}) in the MMG case. Note that for the
\emph{stationary} solutions (\ref{rmet2}) and (\ref{rmet1}), one has the following relation
\begin{equation} 
E_\mathrm{NMG} = \Big( \frac{1+ \omega^2}{2 \omega \ell} \Big) L_\mathrm{NMG} 
\qquad \mbox{and} \qquad
L_\mathrm{MMG} = \Big( \frac{\ell (1+\omega^2)}{2 \omega} \Big) E_\mathrm{MMG}
\label{ELNMMG}
\end{equation}
between the conserved charges. Finally, as a special case of (\ref{rmet2}), the solution (\ref{OTT})
of \cite{Oliva:2009ip} has its charges 
\begin{equation}
E_\mathrm{NMG} = 2 M \,, \quad L_\mathrm{NMG} = 2 M a \,, \qquad \mbox{and} \qquad
E_\mathrm{MMG} = \frac{M a}{\mu \ell^2} \,, \quad L_\mathrm{MMG} = \frac{M}{\mu} \,, 
\label{ELOTT}
\end{equation}
and this is the first time the latter pair has been calculated in the MMG setting.

\begin{table*}[t]\centering
\ra{1.3}
\begin{ruledtabular}
\begin{tabular}
{@{}rcllcll@{}}\toprule
\phantom{abc} & \phantom{abc} & \multicolumn{2}{c}{NMG} & \phantom{abc}& \multicolumn{2}{c}{MMG} \\ 
\cmidrule(lr){3-4} \cmidrule(lr){6-7} 
metric && $E_\mathrm{NMG}$ & $L_\mathrm{NMG}$ && $E_\mathrm{MMG}$ & $L_\mathrm{MMG}$  
\\ \midrule
$ds_\mathrm{BTZ}^2$ \; (\ref{BTZ})& & $2M$ & $2J$ && $\frac{J}{\mu \ell^2}$ & $\frac{M}{\mu}$ \\
$ds_1^2$ \; (\ref{met1})& & $\frac{2 (r_{0} - r_{1}^2)}{\ell^2}$ & $0$ && $0$ & 
``$\frac{(r_{0} - r_{1}^2)}{\mu \ell^2}$" \\
$ds_2^2$ \; (\ref{met2})& & $ -\frac{2 r_{-} r_{+}}{\ell^2}$ & $0$ && $0$ & 
``$-\frac{r_{-} r_{+}}{\mu \ell^2}$" \\
$d\tilde{s}_2^2$ \; (\ref{rmet2}) & & 
$-2 \frac{(2 \alpha^2 -\beta) (1+ \omega^2)}{\ell^2 \omega^2}$ & 
$-4 \frac{(2 \alpha^2 -\beta)}{\omega \ell}$ & & 
$-2 \frac{(2 \alpha^2 - \beta)}{\mu \omega \ell^3}$ & 
$- \frac{(2 \alpha^2 -\beta) (1+ \omega^2)}{\mu \omega^2 \ell^2}$ \\
$d\tilde{s}_1^2$ \; (\ref{rmet1}) & & $-2 \frac{(2 R_0^2 - R_1) (1+ \omega^2)}{\ell^2 \omega^2}$ & 
$-4 \frac{(2 R_0^2 - R_1)}{\omega \ell}$ && 
$-2 \frac{(2 R_0^2 - R_1)}{\mu \omega \ell^3}$ & 
$- \frac{(2 R_0^2 - R_1) (1+ \omega^2)}{\mu \omega^2 \ell^2}$ \\
$ds_\mathrm{OTT}^2$ \; (\ref{OTT})& & $2M$ & $2Ma$ && $\frac{Ma}{\mu \ell^2}$ & $\frac{M}{\mu}$ \\
\bottomrule
\end{tabular}
\end{ruledtabular}
\caption{The energy $E$ and the angular momentum $L$ of the canocical BTZ (\ref{BTZ}), the 
static solutions (\ref{met1}) and (\ref{met2}) and the stationary solutions (\ref{rmet2}), (\ref{rmet1}) 
and (\ref{OTT}) at the respective merger points (\ref{merNMG}) and (\ref{merMMG}) of NMG and 
MMG. Since the charges for GMG are equal to the sum of the corresponding charges
for NMG and MMG, the charges for GMG are not listed here. See (\ref{ELGMG}) for these.}
\label{table}
\end{table*}

As already mentioned, for \emph{all} of these solutions one has 
\begin{equation} 
E_\mathrm{GMG} = E_\mathrm{NMG} + E_\mathrm{MMG} \quad \mbox{and} \quad 
L_\mathrm{GMG} = L_\mathrm{NMG} + L_\mathrm{MMG} \label{ELGMG} \,.
\end{equation} 
I refer the reader to Table \ref{table} for a summary of these calculations. 

\subsection{A peculiarity and a ``duality"}
Out of curiosity, if one is to repeat the calculation described in the paragraph containing 
(\ref{Emet2NMG}) and (\ref{Emet1NMG}) by \emph{bluntly} using the spacelike Killing vector 
\( \bar{\xi}^{\mu} = \big( \partial/\partial \theta \big)^{\mu} \) instead, then one finds, as one should, 
that $L_\mathrm{NMG}=0$ for both solutions in the NMG setting, but that the ``angular 
momenta" of \emph{static} solutions (\ref{met1}) and (\ref{met2}) read
\begin{equation}
L_\mathrm{MMG} = \left\{
\begin{array}{ll}
\frac{(r_{0} - r_{1}^2)}{\mu \ell^2} & \mbox{for} \; (\ref{met1}), \\
-\frac{r_{-} r_{+}}{\mu \ell^2} & \mbox{for} \; (\ref{met2})
\end{array}\right. \label{LmetMMG}
\end{equation}
in the MMG case (and hence the quotation marks in the corresponding entries in Table \ref{table}). 
Note also that if one is to use (\ref{albe}), and respectively (\ref{R0R1}), in (\ref{LrmetMMG}) and 
later to take $\omega \to 0$, then one exactly obtains the expressions given in (\ref{LmetMMG}). 
In retrospect, one has two different examples of \emph{static} solutions which have 
\emph{vanishing energy} and \emph{non-trivial angular momentum} in the MMG setting
at the merger point. Clearly the thermodynamics, if any, of these solutions in the MMG 
setting is anything but ordinary. This is an obvious indication of a serious pathology 
at the merger point of MMG, and may have to do with the observation that the graviton is a
tachyon at the merger point when $\Lambda < 0$ \cite{Bergshoeff:2014pca, Arvanitakis:2014yja}.

Finally, I would like to point out to the following ``duality" between all of these charges:
\begin{equation} 
L_\mathrm{NMG} = 2 \mu \ell^2 \, E_\mathrm{MMG} \qquad \mbox{and} \qquad 
E_\mathrm{NMG} = 2 \mu \, L_\mathrm{MMG} \,. \label{ELrel}
\end{equation} 
I am not aware of the consequences, if any, of this ``duality".

\section{The entropy and the thermodynamics}\label{thermo}
Having come this far and computed the conserved gravitational charges of the static and 
stationary solutions of MMG at its merger point, it would be a shame if one didn't push a bit
further and discuss the thermodynamics of these solutions at least in the MMG setting as 
well. As a minimum requirement, this calls for a convenient tool for determining the entropies 
of these solutions. This in itself introduces a constraint on the solutions whose entropies
can be studied since, as far as I know, the entropy of a given gravitational solution can only 
be reliably studied if it describes a black hole. Hence, I will not consider the static solution 
(\ref{met1}) or its stationary version (\ref{rmet1}) in what follows since they are clearly not black 
holes but solitons.

It turns out that even for the black holes, the calculation of the entropies is quite a nontrivial 
task since, as already mentioned, MMG theory lacks a ``purely" gravitational action that only 
involves the metric and its curvature tensors, and thus is not fit for the application of the more 
or less standard ways of calculating the entropy through e.g. the original Wald formulation. 
Moreover, the presence of the Cotton tensor in the field equations is another source of complication. 
Fortunately, one has the formulation of \cite{Setare:2015nla}, which is especially convenient 
for the MMG case, is inspired by the techniques advanced in \cite{Tachikawa:2006sz} for 
theories that involve Chern-Simons terms in their action (and thus the Cotton tensor in their 
field equations in three dimensions) and relies on the standard tools of a first order formulation 
of MMG. Rather than going over the details of that long calculation here, let me be pragmatic 
and adapt the original formulas given in \cite{Setare:2015nla} to my conventions and give them
\emph{at the merger point of MMG and NMG} only. They then read:
\begin{eqnarray}  
S_\mathrm{MMG} & = & 2 \int_{r=r_h} \frac{d \theta}{\sqrt{g_{\theta\theta}}} \, \Big( 
\bar{\sigma} g_{\theta\theta} + \frac{1}{\mu} \Omega_{\theta\theta} 
+ \frac{\gamma}{\mu^2} S_{\theta\theta} \Big) \,, \label{SMMG} \\
S_\mathrm{NMG} & = & 2 \int_{r=r_h} \frac{d \theta}{\sqrt{g_{\theta\theta}}} \, \Big( 
\sigma g_{\theta\theta} - \frac{1}{m^2} S_{\theta\theta} \Big) \,. \label{SNMG} 
\end{eqnarray}
Here 
\( \Omega^{\mu}_{\;\nu} \equiv \tfrac{1}{2} \epsilon^{\mu\alpha\beta} \, e^{c}_{\;\beta} \,
\nabla_{\nu} e_{c\alpha} \, \)
is the torsion-free dualized spin-connection, $e$ indicates the dreibein and $r_h$ is the
radius of the horizon. Note that for all the solutions considered in this work, 
\( g_{\theta\theta} = r^2 \) and $\theta$ ranges from 0 to $2 \pi$. 

\subsection{The BTZ black hole} 
Let me again use the BTZ metric as a beacon in this endeavor. For the BTZ black hole 
(\ref{BTZ}), \( S_{\theta\theta} = -r^2/(2 \ell^2) \) so that at the merger point (\ref{merNMG}) 
of NMG, one quickly finds 
\[ S_\mathrm{NMG} = 8 \pi R_{+} \,, \] 
where \( R_{+} \) is the larger root of 
\[ \frac{1}{g_{rr}} = -M + \frac{r^2}{\ell^2} + \frac{J^2}{4 r^2} = 0 \qquad \mbox{i.e.} \qquad
R_{\pm} = \frac{\ell}{\sqrt{2}} \sqrt{M \pm \sqrt{M^2 - J^2/\ell^2}} \]
and $R_{+} $ is the location of the event horizon. The BTZ metric, when considered as a solution 
to the cosmological Einstein theory, has entropy \( S = 4 \pi R_{+} \) precisely. Hence, as is 
the case for its energy and angular momentum, its entropy is also doubled when it is 
examined in the NMG setting. For completeness sake, here I rewrite all the relevant 
physical quantities of BTZ in terms of the locations of its horizons $R_{\pm}$:
\begin{eqnarray*}
T & = & \frac{R_{+}^2 - R_{-}^2}{2 \pi \ell^2 R_{+}} \,, \qquad \Omega_{H} = - \frac{g_{t\theta}}{g_{\theta\theta}} \Big|_{r=R_{+}} = \frac{R_{-}}{\ell R_{+}} \,, \\
E_\mathrm{NMG} & = & 2 M = \frac{2}{\ell^2} (R_{+}^2 + R_{-}^2) \,, \quad
L_\mathrm{NMG} = 2 J = \frac{4}{\ell} R_{+} R_{-} \,, \quad
E_\mathrm{NMG} = \frac{S_\mathrm{NMG}^2}{32 \pi^2 \ell^2} 
+ \frac{8 \pi^2 L_\mathrm{NMG}^2}{S_\mathrm{NMG}^2} \,, 
\end{eqnarray*}
where $T$ is the temperature and $\Omega_{H}$ is the angular velocity of the horizon 
computed by standard methods. It immediately follows from these
expressions that the first law of black hole thermodynamics in the form 
\begin{equation}
d E = T \, d S + \Omega_{H} \, d L \label{1stlaw} 
\end{equation} 
does hold at the merger point of NMG.

The only essential difference in the calculation of entropy for the MMG case comes from the 
spin connection. The choice
\[ e^0 = (g_{rr})^{-1/2} dt \,, \quad e^1 = (g_{rr})^{1/2} dr \,, \quad e^2 = r d\theta - \frac{J}{2 r} dt \]
for the dreibein leads to the simple expression \( \Omega_{\theta\theta} = J/2 = R_{+} R_{-}/\ell \).
Using the merger point condition (\ref{merMMG}) of MMG then leads to 
\[ S_\mathrm{MMG} = \frac{4 \pi R_{-}}{\mu \ell} \,. \] 
One now has
\[ E_\mathrm{MMG} = \frac{J}{\mu \ell^2} = \frac{2 R_{+} R_{-}}{\mu \ell^3} \,, \;
L_\mathrm{MMG} = \frac{M}{\mu} = \frac{R_{+}^2 + R_{-}^2}{\mu \ell^2}  \,, \;
E_\mathrm{MMG} = \frac{S_\mathrm{MMG}}{2 \pi \ell} \sqrt{\mu L_\mathrm{MMG} 
- \frac{\mu^2 S_\mathrm{MMG}^2}{16 \pi^2}} \]
analogous to the NMG case and again the first law of black hole thermodynamics (\ref{1stlaw})
holds at the merger point of MMG.

\subsection{The static black hole}
For the static hairy black hole (\ref{met2}), 
\( S_{\theta\theta} = -r (r-r_{-}-r_{+})/(2 \ell^2) \), so that 
\[ S_\mathrm{NMG} = 4 \pi (r_{+} - r_{-}) \]
at the merger point (\ref{merNMG}) of NMG. This expression is indeed identical to what
one would find by using the Wald formulation. Meanwhile, one finds straightforwardly that
\[ T = \frac{1}{4 \pi} \frac{d(1/g_{rr})}{dr}\Big|_{r=r_{+}} = \frac{r_{+} - r_{-}}{4 \pi \ell^2} \,, \] 
and recall that \( E_\mathrm{NMG} = - 2  r_{+} r_{-}/\ell^2 \), so that the first law of 
black hole thermodynamics in the form \( d E = T \, d S \) 
immediately fails unless one considers the ``hairless" limit \( r_{+} + r_{-} \to 0 \).

As for the entropy in the MMG case, since the metric (\ref{met2}) is diagonal,
it follows automatically from \( e^0 = (g_{rr})^{-1/2} dt \,, e^1 = (g_{rr})^{1/2} dr \,, 
e^2 = r d\theta \) that \( \Omega_{\theta\theta} = 0 \). These lead to
\[ S_\mathrm{MMG} = 4 \pi \bar{\sigma} (r_{+} + r_{-}) \] 
at the merger point of MMG. Since I already found \( E_\mathrm{MMG} = 0 \) for (\ref{met2}), 
\( d E = T \, d S \) doesn't hold unless one considers the ``hairless" limit again.

By now the lesson is clear: Unless one goes to the ``hairless" limit, there is no reason 
to expect the first law of black hole thermodynamics to hold for a black hole with hair. However,
it is instructive to see how (\ref{SMMG}) and (\ref{SNMG}) can be used. Thus, let me continue
pedantically.

\subsection{The stationary black hole} 
For the ensuing discussion, recall the choices made earlier for the existence of a 
horizon (\ref{choices}) so that the stationary metric (\ref{rmet2}) can be interpreted as 
a black hole. The following dreibein choice
\begin{eqnarray*}
e^0 & = & \frac{\sqrt{f(r) \big( f(r) - (1-\omega^2) r^2 \big)}}{r \omega \ell} dt \,, \quad 
e^1 =  \frac{r \omega \ell}{\sqrt{f(r) - (1-\omega^2) r^2}} \,
\frac{1}{\sqrt{r^2 + \alpha^2 -\beta}} dr \,, \\
e^2 & = & \frac{f(r) - r^2}{r \omega \ell} dt + r d\theta \,,
\end{eqnarray*}
that naturally follows from the alternative form (\ref{altrmet2}), leads to
\begin{eqnarray} 
\Omega_{\theta\theta} & = & 
\frac{\big( r f^{\prime}(r) - 2 f(r) \big) \sqrt{r^2 + \alpha^2 -\beta}}{2 \ell \omega \sqrt{f(r)}} \,,
\quad \mbox{and} \quad 
\Omega_{\theta\theta}\Big|_{r=r_{h}} = \frac{\beta + \alpha r_{h} \sqrt{1-\omega^2}}{\ell \omega} \,, \\
S_{\theta\theta} & = & - \frac{r^2}{2 \ell^2} + \frac{\alpha}{\ell^2 \omega^2} \sqrt{f(r)} \,,
\quad \mbox{and} \quad 
S_{\theta\theta}\Big|_{r=r_{h}} =  \frac{r_{h}^2}{2 \ell^2} 
\left( -1 + \frac{2 \alpha \, \sqrt{1-\omega^2}}{r_{h} \, \omega^2} \right) \,.
\end{eqnarray}
Using these in (\ref{SNMG}) and (\ref{SMMG}), one finds 
\begin{eqnarray}
S_\mathrm{NMG} & = & \frac{8 \pi}{\omega^2} \sqrt{\alpha^2(1-\omega^2) + \beta \omega^2} \,, \\
S_\mathrm{MMG} & = & \frac{4 \pi}{\mu \ell \omega} \left( \frac{\beta}{r_{h}} 
+ \alpha \, \sqrt{1-\omega^2} \Big( 1+ \frac{2 \mu \ell \bar{\sigma}}{\omega} \Big) \right) 
= \frac{8 \pi \alpha \bar{\sigma}}{\omega^2} \sqrt{1-\omega^2} + \frac{\omega}{2 \mu \ell} S_\mathrm{NMG} \,. \label{smmg}
\end{eqnarray}
Meanwhile the angular velocity and the temperature of the horizon are given by
\begin{eqnarray}
\Omega_{H} & = & - \frac{g_{t\theta}}{g_{\theta\theta}} \Big|_{r=r_{h}} = \frac{\omega}{\ell} \,, \\
T & = &  \frac{\sqrt{1-\omega^2} \, \sqrt{r_{h}^2 + \alpha^2 - \beta}}{2 \pi \omega^2 \ell^2} 
\left( \omega^2 - \frac{\alpha}{\sqrt{r_{h}^2 + \alpha^2 - \beta}} \right)
= \frac{(1-\omega^2)}{2 \pi \omega^2 \ell^2} \sqrt{\alpha^2(1-\omega^2) + \beta \omega^2} \,.
\end{eqnarray}
As already advertised, it is a simple exercise to check from these expressions, (\ref{ErmetMMG}) 
and (\ref{LrmetMMG}) that the first law of black hole thermodynamics in the form  (\ref{1stlaw}) 
does not hold unless one goes to the ``hairless" limit \( \alpha \to 0 \), i.e. the BTZ case.

Finally, the expressions of the entropy, the angular velocity and the temperature of the horizon
given above can be translated to find the corresponding quantities of
$ds_\mathrm{OTT}^2$ (\ref{OTT}) using (\ref{ott}). It is gratifying to see that these indeed 
reduce to their respective counterparts in the NMG setting already given in 
\cite{Giribet:2009qz}\footnote{The hair parameter $b$ in (\ref{ott}) is related to the one used in
\cite{Giribet:2009qz} via \( b \to b \, \Xi^{1/2} \), and I use the $b$ of \cite{Giribet:2009qz} in 
what follows.}. For the sake of completeness, these thermodynamical quantities read
\begin{eqnarray}
\Omega_{H} & = & \frac{a}{\ell^2 (1+ \Xi^{1/2})} \,, \\
T & = & \frac{\sqrt{2}}{4 \pi \ell} \sqrt{ \big(1+ \Xi^{1/2} \big) \Big(M + \frac{b^2 \ell^2}{4} \Big)} \,, \\
S_\mathrm{NMG} & = & 4 \sqrt{2} \, \pi \ell  \sqrt{ \big(1+ \Xi^{1/2} \big) \Big(M + \frac{b^2 \ell^2}{4} \Big)} \,, \\
S_\mathrm{MMG} & = & 2 \sqrt{2} \, \pi \sqrt{1+ \Xi^{1/2}} \left( - b \ell^2 \bar{\sigma} 
+\frac{\ell}{a \mu} \big(1-\Xi^{1/2} \big) \sqrt{M + \frac{b^2 \ell^2}{4}} \right) \,. \label{smmgott}
\end{eqnarray}

\section{Discussion}\label{disc}
In this work, first I found all the static circularly symmetric solutions of MMG at its merger point.
It was a surprise to find that the metric (\ref{met1}), which was already known to be a solution 
of NMG and describes a gravitational soliton, was overlooked in the literature. Together
with the static hairy black hole (\ref{met2}), I was then able to boost these two to find their 
stationary versions, which are themselves circularly symmetric and ``new" solutions when 
considered in the MMG setting. I have shown that the stationary black hole (\ref{rmet2}) is 
indeed the already known black hole (\ref{OTT}) in disguise, thanks to the identifications 
(\ref{ott}). However the derivation presented here should put (\ref{OTT}) in a more 
comprehensible ground, at least for the sake of novices in the field, and (\ref{rmet2}) is 
certainly friendlier to work with. The rotating soliton (\ref{rmet1}) is also new, and considering
the result I found in the penultimate paragraph of subsection \ref{scss}, \emph{all} of the 
metrics studied in this work, \emph{apart} from the Lifshitz ones, are solutions to NMG, MMG and 
GMG at their respective merger points. Provided that both NMG and MMG admit CFTs at
the boundary, this ``duality" may point out to a certain relationship between these two
CFTs.

I then calculated the conserved gravitational charges of the asymptotically AdS solutions,
and made the observation that these charges obey the ``duality" relation (\ref{ELrel}). Even
though I am not aware of any particular meaning this might have, it is interesting that this
essentially swaps the roles of energy and angular momentum as one changes the setting
from NMG to MMG, and vice versa, in which one is examining a given solution. I was also
able to calculate the entropies, and other relevant thermodynamical quantities, of the
asymptotically AdS black holes, and showed that the first law of black hole thermodynamics
does not hold unless one considers the ``hairless" limit. In \cite{Giribet:2009qz}, it is argued
for (\ref{OTT}) that the variations (in the thermodynamical sense) in the hair parameter 
can be reabsorbed by measuring the gravitational charges not with respect to the AdS 
background, but with respect to an ``extremal" background which itself has its energy given 
entirely in terms of the hair parameter (see \cite{Giribet:2009qz} for details). The same 
``trick" can certainly be played here to restore the first law of black hole thermodynamics 
in the form \( d (E - E_{h}) = T \, d S + \Omega_{H} \, d (L - L_{h}) \,, \) where $E_{h}$ 
and $L_{h}$ indicate the energy and the angular momentum of the ``extremal" background. 
It should certainly be worth the effort to understand the microscopic origins of the entropy
$S_\mathrm{MMG}$ (\ref{smmg}) of (\ref{rmet2}) (and thus (\ref{smmgott}) of (\ref{OTT})) 
and to study it from the CFT side via a Cardy-like formula.

\begin{acknowledgments}
I would like to thank N. Sad{\i}k De\u{g}er for collaboration on the early stages of this work.
\end{acknowledgments}

\appendix
\section{\label{appa}}
Here I explicitly show the steps to take when one applies the ``boosting" trick
that lets one to obtain (\ref{rmet2}) from (\ref{met2}). To this end, let me start 
by rewriting (\ref{met2}) as (setting $\Lambda = -1/\ell^2 < 0$)
\begin{equation} 
ds^2 = - u(\rho) \, d\tau^2 + \frac{d\rho^2}{u(\rho)} + \rho^2 \, d\phi^2 \,, \qquad
u(\rho) \equiv \frac{(\rho-r_{-}) (\rho-r_{+})}{\ell^2} \,, \label{bas}
\end{equation}
by an obvious renaming of the coordinates and the metric functions.
Assuming \( |\omega|<1 \), apply the linear coordinate transformation
\[ \left[ 
\begin{array}{c}
d\tau \\
d\phi
\end{array}
\right] =
\frac{1}{\sqrt{1-\omega^2}}\left[
\begin{array}{lr}
  1 & -\omega \ell \\
  -\omega/\ell & 1 
\end{array}
\right]
\left[ 
\begin{array}{c}
d t \\
d \theta
\end{array}
\right] \]
to (\ref{bas}) to arrive at
\begin{equation}
ds^2 = \big( \frac{\omega^2}{\ell^2} \rho^2 - u(\rho) \big) \, \frac{dt^2}{(1-\omega^2)} 
- \frac{2 \omega \ell}{(1-\omega^2)} \, \big( \frac{\rho^2}{\ell^2} - u(\rho) \big) dt \, d\theta
+ \frac{d\rho^2}{u(\rho)} 
+ \big( \rho^2 - \omega^2 \, \ell^2 \, u(\rho) \big)\, \frac{d\theta^2}{(1-\omega^2)} \,.
\label{ara}
\end{equation}
Using the explicit form of $u(\rho)$ and a coordinate redefinition 
\( r^2 \equiv \rho^2 + 2 \alpha \rho + \beta \), the metric (\ref{ara}) can be cast in the form
(\ref{rmet2}) in terms of the parameters $\alpha$ and $\beta$ defined in (\ref{albe}).


\begin{thebibliography}{99}

\bibitem{Deser:1982vy} 
  S.~Deser, R.~Jackiw and S.~Templeton,
  ``Three-Dimensional Massive Gauge Theories,"
  Phys.\ Rev.\ Lett.\  {\bf 48}, 975 (1982).

  S.~Deser, R.~Jackiw and S.~Templeton,
  ``Topologically Massive Gauge Theories,"
  Annals Phys.\  {\bf 140}, 372 (1982)
  [Erratum-ibid.\  {\bf 185}, 406 (1988)]
  [Annals Phys.\  {\bf 185}, 406 (1988)]
  [Annals Phys.\  {\bf 281}, 409 (2000)].

\bibitem{Bergshoeff:2009hq} 
  E.~A.~Bergshoeff, O.~Hohm and P.~K.~Townsend,
  ``Massive Gravity in Three Dimensions,''
  Phys.\ Rev.\ Lett.\  {\bf 102}, 201301 (2009)
  [\href{http://arxiv.org/abs/0901.1766}{{\tt arXiv:0901.1766}}].
  
  E.~A.~Bergshoeff, O.~Hohm and P.~K.~Townsend,
  ``More on Massive 3D Gravity,''
  Phys.\ Rev.\ D {\bf 79}, 124042 (2009)
  [\href{http://arxiv.org/abs/0905.1259}{{\tt arXiv:0905.1259}}].
  
\bibitem{Bergshoeff:2014pca} 
  E.~Bergshoeff, O.~Hohm, W.~Merbis, A.~J.~Routh and P.~K.~Townsend,
  ``Minimal Massive 3D Gravity,''
  Class.\ Quant.\ Grav.\  {\bf 31}, 145008 (2014)
  [\href{http://arxiv.org/abs/1404.2867}{{\tt arXiv:1404.2867}}].

\bibitem{Arvanitakis:2014xna} 
  A.~S.~Arvanitakis and P.~K.~Townsend,
  ``Minimal Massive 3D Gravity Unitarity Redux,''
  Class.\ Quant.\ Grav.\  {\bf 32}, no. 8, 085003 (2015)
  [\href{http://arxiv.org/abs/1411.1970}{{\tt arXiv:1411.1970}}].

\bibitem{Arvanitakis:2014yja} 
  A.~S.~Arvanitakis, A.~J.~Routh and P.~K.~Townsend,
  ``Matter coupling in 3D `minimal massive gravity',''
  Class.\ Quant.\ Grav.\  {\bf 31}, no. 23, 235012 (2014)
  [\href{http://arxiv.org/abs/1407.1264}{{\tt arXiv:1407.1264}}].

\bibitem{Banados:1992wn} 
  M.~Banados, C.~Teitelboim and J.~Zanelli,
  ``The black hole in three-dimensional space-time,''
  Phys.\ Rev.\ Lett.\  {\bf 69}, 1849 (1992)
  [\href{http://arxiv.org/abs/hep-th/9204099}{{\tt arXiv:hep-th/9204099}}].

\bibitem{Alishahiha:2014dma} 
  M.~Alishahiha, M.~M.~Qaemmaqami, A.~Naseh and A.~Shirzad,
  ``On 3D Minimal Massive Gravity,''
  JHEP {\bf 1412}, 033 (2014)
  [\href{http://arxiv.org/abs/1409.6146}{{\tt arXiv:1409.6146}}].
  
\bibitem{Giribet:2014wla} 
  G.~Giribet and Y.~V{\'a}squez,
  ``Minimal Log Gravity,''
  Phys.\ Rev.\ D {\bf 91}, no. 2, 024026 (2015)
  [\href{http://arxiv.org/abs/1411.6957}{{\tt arXiv:1411.6957}}].
  
\bibitem{Arvanitakis:2015yya} 
  A.~S.~Arvanitakis,
  ``On Solutions of Minimal Massive 3D Gravity,''
  Class.\ Quant.\ Grav.\  {\bf 32}, no. 11, 115010 (2015)
   [\href{http://arxiv.org/abs/1501.01808}{{\tt arXiv:1501.01808}}].

\bibitem{Deger:2015wpa} 
  N.~S.~Deger and {\"O}. Sar{\i}o\u{g}lu,
  ``Kundt solutions of Minimal Massive 3D Gravity,''
  Phys.\ Rev.\ D {\bf 92}, no. 10, 104015 (2015)
  [\href{http://arxiv.org/abs/1505.03387}{{\tt arXiv:1505.03387}}].
  
\bibitem{Charyyev:2017uuu} 
  J.~Charyyev and N.~S.~Deger,
  ``Homogeneous Solutions of Minimal Massive 3D Gravity,''
  Phys.\ Rev.\ D {\bf 96}, no. 2, 026024 (2017)
  [\href{http://arxiv.org/abs/1703.06871}{{\tt arXiv:1703.06871}}]. 

\bibitem{Afshar:2014ffa} 
  H.~R.~Afshar, E.~A.~Bergshoeff and W.~Merbis,
  ``Extended massive gravity in three dimensions,''
  JHEP {\bf 1408}, 115 (2014)
  [\href{http://arxiv.org/abs/1405.6213}{{\tt arXiv:1405.6213}}].
  
  B.~Tekin,
  ``Bulk and boundary unitary gravity in 3D: MMG$_2$,''
  Phys.\ Rev.\ D {\bf 92}, no. 2, 024008 (2015)
  [\href{http://arxiv.org/abs/1503.07488}{{\tt arXiv:1503.07488}}].

\bibitem{Kachru:2008yh} 
  S.~Kachru, X.~Liu and M.~Mulligan,
  ``Gravity duals of Lifshitz-like fixed points,''
  Phys.\ Rev.\ D {\bf 78}, 106005 (2008)
  [\href{http://arxiv.org/abs/hep-th/0808.1725}{{\tt arXiv:hep-th/0808.1725}}].

\bibitem{Oliva:2009ip} 
  J.~Oliva, D.~Tempo and R.~Troncoso,
  ``Three-dimensional black holes, gravitational solitons, kinks and wormholes for BHT massive gravity,''
  JHEP {\bf 0907}, 011 (2009)
  [\href{http://arxiv.org/abs/0905.1545}{{\tt arXiv:0905.1545}}].  
  
 \bibitem{Perez:2011qp} 
  A.~Perez, D.~Tempo and R.~Troncoso,
  ``Gravitational solitons, hairy black holes and phase transitions in BHT massive gravity,''
  JHEP {\bf 1107}, 093 (2011)
   [\href{http://arxiv.org/abs/1106.4849}{{\tt arXiv:1106.4849}}].  

\bibitem{Martinez:1999qi} 
  C.~Martinez, C.~Teitelboim and J.~Zanelli,
  ``Charged rotating black hole in three space-time dimensions,''
  Phys.\ Rev.\ D {\bf 61}, 104013 (2000)
  [\href{https://arxiv.org/abs/hep-th/9912259}{{\tt arXiv:hep-th/9912259}}].

\bibitem{Deser:2002jk} 
  S.~Deser and B.~Tekin,
  ``Energy in generic higher curvature gravity theories,''
  Phys.\ Rev.\ D {\bf 67}, 084009 (2003)
  [\href{http://arxiv.org/abs/hep-th/0212292}{{\tt arXiv:hep-th/0212292}}].
  
  S.~Deser and B.~Tekin,
  ``Energy in topologically massive gravity,''
  Class.\ Quant.\ Grav.\  {\bf 20}, L259 (2003)
   [\href{http://arxiv.org/abs/gr-qc/0307073}{{\tt arXiv:gr-qc/0307073}}].
  
   B.~Tekin,
  ``Minimal Massive Gravity: Conserved Charges, Excitations and the Chiral Gravity Limit,''
  Phys.\ Rev.\ D {\bf 90}, no. 8, 081701 (2014)
  [\href{http://arxiv.org/abs/1409.5358}{{\tt arXiv:1409.5358}}].
  
\bibitem{Setare:2015nla} 
  M.~R.~Setare and H.~Adami,
  ``Black hole entropy in the Chern–Simons-like theories of gravity and Lorentz-diffeomorphism Noether charge,''
  Nucl.\ Phys.\ B {\bf 902}, 115 (2016)
  [\href{http://arxiv.org/abs/1509.05972}{{\tt arXiv:1509.05972}}]. 

\bibitem{Tachikawa:2006sz} 
  Y.~Tachikawa,
  ``Black hole entropy in the presence of Chern-Simons terms,''
  Class.\ Quant.\ Grav.\  {\bf 24}, 737 (2007)
   [\href{http://arxiv.org/abs/hep-th/0611141}{{\tt arXiv:hep-th/0611141}}].
   
\bibitem{Giribet:2009qz} 
  G.~Giribet, J.~Oliva, D.~Tempo and R.~Troncoso,
  ``Microscopic entropy of the three-dimensional rotating black hole of BHT massive gravity,''
  Phys.\ Rev.\ D {\bf 80}, 124046 (2009)
  [\href{http://arxiv.org/abs/0909.2564}{{\tt arXiv:0909.2564}}].  

\end{thebibliography}
\end{document}